\documentclass[preprint,showpacs,preprintnumbers,amsmath,amssymb]{revtex4}

\usepackage{epsfig}% Include figure files
\usepackage{dcolumn}% Align table columns on decimal point
\usepackage{bm}% bold math
\usepackage{float} 
\begin{document}

\title{The encapsulation of hydrophobic drugs in Pluronic F127 micelles: the effects of drug hydrophobicity, solution temperature and pH}

\author{Rajib Basak}
\email{rajib@rri.res.in}
\affiliation{Raman Research Institute, Bangalore 560080, INDIA}
\author{Ranjini Bandyopadhyay}
\email{ranjini@rri.res.in}
\affiliation{Raman Research Institute, Bangalore 560080, INDIA}
\vspace{0.5cm}
%$^{1}$ Raman Research Institute, Bangalore 560080, INDIA\\
%$^{2}$ Department of Chemical Engineering, Indian Institute of Technology Kanpur,
%Kanpur 208016, INDIA\\
%\affiliation{

\date{\today}% It is always \today, today,


\newpage
%\centerline{Summary}
\vspace{1cm}
\begin{abstract}

 Three drugs, Ibuprofen, Aspirin and Erythromycin, are encapsulated in spherical Pluronic F127 micelles. The shapes and the size distributions of the micelles in dilute, aqueous solutions, with and without drugs, are ascertained using cryo- Scanning Electron Microscopy and Dynamic Light Scattering (DLS) experiments, respectively. Uptake of drugs above a threshold concentration is seen to reduce the critical micellization temperature of the solution. The mean hydrodynamic radii and polydispersities of the micelles are found to increase with decrease in temperature and in the presence of drug molecules. The hydration of the micellar core at lower temperatures is verified using fluorescence measurements. Increasing solution pH leads to the ionization of the drugs incorporated in the micellar cores.  This causes rupture of the micelles and release of the drugs into the solution at the highest solution pH value of 11.36 investigated here and is studied using DLS and fluorescence spectrocopy.
%Valid PACS numbers may be entered using the \verb+\pacs{#1}+ command.
\end{abstract}
\maketitle
\renewcommand{\thesection}{\arabic{section}}

\section{Introduction}
In recent times, block copolymers have emerged as a potential agent for targeted drug delivery and gene therapy \cite{Batrakova_jcc,kwon_addr,kaki_addr, scher_ijp,kabanov_addr}. One such block copolymer proposed for controlled drug delivery is Pluronic, which has a triblock  PEO-PPO-PEO  structure (PEO: polyethylene oxide, PPO: polypropylene oxide). At high temperatures, the central PPO block becomes hydrophobic, while the PEO blocks remain hydrophilic \cite{jacob_pnas}. Due to this amphiphilic nature, Pluronic molecules, above a critical temperature and concentration, self-aggregate in aqueous solutions to form spherical micelles with hydrophobic PPO cores surrounded by hydrophilic PEO coronas. As the copolymer concentration is increased, the micelles can arrange themselves in cubic crystalline order  \cite{yardimci_jcp,mortensen_prl}. This results in a transition from a liquid to a soft solid phase \cite{harsha_pre}. The micellization process and the phase behavior of pure Pluronic solutions are found to depend upon solution temperature, copolymer concentration and molecular architecture and have been studied using differential scanning calorimetry (DSC), static and dynamic light scattering (SLS and DLS, respectively), small angle neutron scattering (SANS) and surface tension measurements \cite{wanka_macro,alex_lang,mortensen_macro,prud_lang}.\\
\indent The sizes of Pluronic micelles and the phase behavior of Pluronic solutions show dramatic changes in the presence of certain additives. The addition of the anionic surfactant sodium dodecyl sulphate (SDS) to micellar Pluronic solutions, for example, leads to the formation of mixed micelles. These solutions have been studied using DSC, isothermal titration calorimetry (ITC), SLS and DLS, electromotive force (emf) and surafce tension measurements and rheology \cite{hecht_lang,desai_jcis,li_lang,jansson_jpcb,hecht_jpc,thurn_lang, basak_epje}.\\
  \indent The formation of micellar block copolymer-drug complexes was first proposed by Dorn \textit{et al.} \cite{ring_book}. When hydrophobic drug molecules are mixed with suitable quantities of Pluronic molecules and the temperature is raised, the drug molecules accumulate in the hydrophobic PPO cores. The hydrated PEO coronas are non-toxic and prevent the drug molecules from being removed from the core. The solubilities of the hydrophobic drugs therefore increase substantially in an aqueous medium, enhancing the bioavailability of the drugs \cite{torch_jcr}. It has been reported that the passive accumulation of drugs encapsulated in Pluronic micelles at solid tumor cells is more efficient than that of free drugs \cite{kwon_addr}. This arises from the long circulation time of the drug-encapsulated micelles and the slow dissociation of drugs from these micelles in the blood circulation system \cite{kwon_addr}. Drug-encapsulated Pluronic micelles can also enhance the transport of drugs across the blood-brain and intestinal barriers \cite{Batrakova_jcc,kwon_addr}. These and other pharmaceutical advantages of Pluronic micelles make it a serious contender as a drug carrier  \cite{moghimi_nanotech}. \\
 \indent The solubilization of drugs in Pluronic micelles has been extensively studied in recent times. It is reported in the literature that the presence of the hydrophobic drug molecules naproxen and indomethacin in F127 solutions results in slight decreases in the micellar sizes and aggregation numbers, in addition to a lowering of the gelation temperature \cite{sharma_ijp}. The aggregation behavior of F127 micelles was studied systematicaly when drugs of varying hydrophobicities were incorporated in the micellar cores \cite{sharma_collsurf}. It was found that the most hydrophobic drugs increased the sizes of the micellar core and corona while generally decreasing the micellar aggregation numbers. Scherlund \textit{et al.} observed a decrease in the critical micellization temperature (CMT) and the gelation temperature when local anesthetics were added to Pluronic solutions \cite{scher_ijp}. These authors further confirmed  that these temperatures decreased with dilution and increase in solution pH. More recently, Foster \textit{et al.} encapsulated Ibuprofen molecules in Pluronic P-104 and P-105 solutions and Flurbiprofen molecules in P103 and P123 solutions. Using SANS and pulsed-field gradient stimulated-echo nuclear magnetic resonance (NMR) measurements, they demonstrated that the encapsulation of drugs favors the micellization process, resulting in an increase in the aggregation number and the micellar core radius \cite{foster_lang,alexander_lang2}. A strong dependence of the aggregation number and the core radius on the solution pH was reported when Ibuprofen was added to P104 solutions and Flurbiprofen to P103 and P123 solutions \cite{foster_lang2,alexander_lang}.\\
  \indent In this work, the non-steroidal anti-inflammatory drug Ibuprofen, the salicylate analgesic drug Aspirin and the macrolide antibiotic drug Erythromycin are encapsulated in the hydrophobic cores of spherical Pluronic F127 micelles. The dilute, completely fluid-like F127 solutions used here are studied at several temperatures and pH values and are prepared at concentrations that are much higher than the critical micellization concentrations of F127 at the temperatures studied. The encapsulation of drugs in Pluronic F127 micelles is studied using cryo-SEM, DLS, small angle X-ray scattering and fluorescence spectroscopy. The changes in the CMTs of the solutions,  the variations in the  hydrodynamic radii of the micelles containing the encapsulated drugs and the micellar polydispersities are investigated using DLS. The temperature dependence of the encapsulation process and the penetration of solvent in the micellar core are studied using fluorescence spectroscopy. Finally, the release of the drugs from the micellar cores when solution pH is increased is reported.

\section{ MATERIALS AND METHODS}
\renewcommand{\thesubsection}{\thesection.\arabic{subsection}}
\indent \textbf{2.1. Sample Preparation.} Pluronic F127, Ibuprofen, Aspirin and Erythromycin were purchased from Sigma-Aldrich and used as received without further purification. Pluronic F127  is a triblock copolymer [PEO$_{100}$PPO$_{70}$PEO$_{100}$] and has a molecular weight of 12,600 g/mol, a critical micellization concentration of 0.725 wt\% at 25$^{\circ}$C. Ibuprofen (mol. wt. 206.29 g/mol), Aspirin (mol. wt. 180.16 g/mol) and Erythromycin (mol. wt. 733.93 g/mol) are three aromatic hydrophobic drug molecules characterized by different chemical architectures (Fig. 1). All these drug molecules are weakly acidic in aqueous solutions. Aspirin is more acidic (pKa 3.5) compared to Ibuprofen (pKa 5.38) and Erythromycin (pKa 8.8) \cite{mitchell_jps,domanska_jpcb,gerzon_jacs}. To prepare a pure F127 solution, an appropriate amount of F127 was dissolved in deionized and distilled Millipore water and stirred vigorously with a magnetic stirrer. Of the three drug molecules used in this study, Ibuprofen is the most hydrophobic, followed by Erythromycin and Aspirin (Table 1). 

 F127 solutions were prepared at concentrations 5.26 wt\%, 2.04 wt\% and 1.02 wt\%. These concentrations are much higher than the CMCs of the F127 solutions at the temperatures studied in this work. Drug molecules were incorporated  in Pluronic F127 micelles by vigorously stirring aqueous mixtures of the drugs and F127 in an ultrasonicator at temperatures between $40^{\circ}$C and $60^{\circ}$C.\\ 

 \indent \textbf{2.2. DLS Measurements.} A BIC (Brookhaven Instruments Corporation) BI-200SM spectrometer was used to measure the scattered light intensity at scattering angles between $60^{\circ}$ and $135^{\circ}$.  A 150 mW solid state laser (NdYVO$_{4}$, Spectra Physics Excelsior) with wavelength 532 nm was used as the light source. The sample cell was held in a brass thermostat block filled with decalin, a refractive index matching liquid. The temperature of the sample cell was controlled between 10$^{\circ}$C and 80$^{\circ}$C with the help of a temperature controller (Polyscience Digital). A Brookhaven BI-9000AT Digital Autocorrelator was used to measure the intensity auto-correlation functions of the light scattered from the samples. The intensity auto-correlation function $G^{(2)}(\tau)$ is defined as $G^{(2)}(\tau) = \frac{<I(0)I(\tau)>}{<I(0)>^{2}} =  1+ A|g^{(1)}(\tau)|^{2}$ \cite{bern_pecora},
where $I(\tau)$ is the intensity at time $\tau$, $g^{(1)}(\tau)$ is the normalized electric field auto-correlation function, A is the coherence  factor, and the angular brackets $< >$ represent an average over time. If the sample is monodisperse and dilute,  $g^{(1)}(\tau) \sim \exp(-\tau/\tau_{R})$, where $\tau_{R}$ is the relaxation time. For spherical colloidal particles diffusing in a solvent of refractive index $n$, $1/\tau_{R} = Dq^{2}$, where $D$ is the translational diffusion coefficient and $q={\frac{4\pi{n}}{\lambda}}\sin (\theta/2)$ is the scattering wave vector at a scattering angle $\theta$ and for a wavelength $\lambda$. The effective hydrodynamic radius $R_{H}$ of the scatterer is estimated using the Stokes-Einstein relation $ D = k_{B}T/6\pi\eta{R_{H}}$, where $k_{B}$ is the Boltzmann constant, $T$ is the temperature and $\eta$ is the viscosity of the solvent \cite{einstein_rad}. Since the samples used in this study were highly polydisperse, the normalized correlation functions [$ C(\tau)$ = ${G^{(2)}}$$(\tau)\ -1$] were fitted to stretched exponential functions of the form $C(\tau) \sim \exp(-\tau/\tau_{R})^{\beta}$. Polydispersity values were extracted using the protocol detailed in Part B of the `Analysis and Calculations' section of Supporting Information. 

\indent \textbf{2.3. Cryo-SEM (scanning electron microscopy) measurements.} For the cryo-SEM experiments, the samples were cryo-fractured by quickly dipping them into liquid nitrogen. The frozen samples were then transferred to a PP3000T cryo unit (Quorum Technologies) and cut with a cold knife. The fractured and cut samples were sublimated at $-130^{\circ}$C for 5 minutes and then sputtered with platinum for ninety seconds inside the unit. The samples were imaged using a Zeiss Ultra Plus cryo-SEM setup at a temperature of $-160^{\circ}$C.

\indent \textbf{2.4. Fluorescence measurements.} For each fluorescence measurement, $5\times 10^{-7}$M pyrene was added to the Pluronic micellar solutions. The samples were stirred in an ultrasonicator for sufficient time to encapsulate the pyrene molecules in the micellar cores. The samples were then kept in quartz cells and excited with light of wavelength 339 nm in a Horiba Jobin Yvon FluoroMax-4 Spectrofluorometer. Changes in the emission spectra due to drug incorporation and changes in temperature and pH were detected using  a R928P photomultiplier tube. This method is widely used to study micellization phenomena \cite{wilhelm_macro,yekta_macro,asta_macro,zana_macro,naka_1,naka_2}.

\indent \textbf{2.5. Small angle X-ray scattering (SAXS).} Small angle X-ray scattering (SAXS) was performed using a Hecus S3-Micro System using a Cu $K_{\alpha} (\lambda = 1.54{\AA} )$. The scattered intensity was measured with a one-dimensional position sensitive detector (PSD). The X-ray diffraction studies were carried out on samples loaded in 1.0 mm diameter glass capillaries.

\indent \textbf{2.6. pH measurement.} pH measurements were done using an Eutech pH-meter. The pH-meter was calibrated by using three buffer solutions of pH 4.01, 7.00 and 10.01.

\section{Results and Discussion}

  Direct visualization of the Pluronic F127 micelles was performed with cryo-SEM. The samples, initially at room temperature ($25^{\circ}$C), were cryo-fractured in liquid nitrogen. A representative image of a 5.26 wt\%  F127 solution is shown in Fig. 2(a). The globular structures of F127 micelles, of average sizes  68-70 nm, are clearly visible. When drug molecules are added Pluronic micellar solutions, they assemble in the hydrophobic core. The cryo-SEM images of F127 micelles encapsulating 0.1 wt\% each of Ibuprofen, Aspirin and Erythromycin respectively are shown in Figs. 2 (b), (c) and (d), respectively. In each sample, the micelles retain their globular structures. In the concentration range investigated, cylindrical or lamellar aggregates \cite{nag_csb, guo_pb} were not observed. The size distributions of these drug-encapsulated micelles are plotted in Fig. S1 of Supporting Information and the methods adopted for these estimates are detailed in Part A of the `Analysis and Calculations' section.

\indent In contrast to cryo-SEM, DLS is a non-invasive technique and yields sizes and distributions over the entire scattering volume. Systematic DLS measurements were next performed at several temperatures with 5.26 wt\% F127 solutions, with and without drugs. In these measurements, the normalized intensity auto-correlation functions $ C(\tau)$ were measured at sixl scattering angles while varying the delay times $\tau$ at several temperatures within the range 12$^{\circ}$C - 60$^{\circ}$C. Fig. 3 shows the data at 60$^{\circ}$C (squares), 40$^{\circ}$C (down-triangles) and 25$^{\circ}$C (up-triangles) for 5.26 wt\% F127 solutions. The fits to the stretched exponential functional form, 
$C(\tau) \sim \exp(-\tau/\tau_{R})^{\beta}$, are shown by solid lines. For every sample, $ C(\tau)$ is measured at six different $q$ values and  the corresponding $\tau_{R}$ and $\beta$ are extracted. The mean relaxation time $<\tau_{R}>$ is estimated using the relation $<\tau_{R}>$ = $\tau_{R}/\beta*\Gamma({1/\beta})$ (details of the calculation are included in Part B of the `Analysis and Calculations' section of  Supporting Information). The inset of Fig. 3 shows that at all three temperatures, $1/<\tau_{R}>$ varies linearly with $q^{2}$ with $1/<\tau_{R}>$ $\rightarrow 0$ as $q$ $\rightarrow 0$, a  signature of the diffusive motion of the nearly spherical micelles. If the sample temperature is lowered to $20^{\circ}$C (circles in Fig. 3), $C(\tau)$ no longer fits to a stretched exponential form. This indicates the breakup of the spherical micelles into free unimers in solution and indirectly establishes that the CMT of the sample lies between $20^{\circ}$C and 25$^{\circ}$C. By systematically inspecting the shape change of $C(\tau)$ from stretched exponential to non-exponential as temperature is decreased, the CMT of this sample is estimated to be $23^{\circ}\pm 1^{\circ}$C. This estimate matches well with previous results obtained using DSC \cite{wanka_macro}, SLS and fluorescence spectroscopy \cite{bohorquez_jcis} and is also confirmed by our cryo-SEM images which show that the F127 solution, quenched rapidly from 25$^{\circ}$C, is constituted by spherical micelles (Fig. 2(a)).\\

\indent  DLS measurements were next performed after incorporating different quantities of drug molecules in 5.26 wt\% F127  solutions in the temperature range $12^{\circ}$C - $60^{\circ}$C. It was observed that the $C(\tau)$ data in this entire temperature range fit well to stretched exponential  functions (Fig. 4). For all the data acquired, $1/<\tau_{R}>$ varies linearly with $q^2$ (inset (a) of Fig. 4), suggesting that the samples are in the micellar liquid phase. The CMTs  of all the drug-incorporated micellar samples therefore lie below the experimental temperature limit of $12^{\circ}$C. The reduction in the CMTs of the micellar solutions observed here due to drug encapsulation is consistent with previous reports \cite{scher_ijp,foster_lang}. The hydrophobicity of the PPO cores increases considerably due to the addition of the drug molecules. This favours micellar aggregation at lower temperatures, resulting in the observed lowering of the CMT.

 The acceleration of the aggregation process with increasing hydrophobicity of the micellar core  is further confirmed by our observation that increasing drug hydrophobicity leads to micelle formation at lower threshold concentrations. The shapes of the $C(\tau$) \textit{vs.} $\tau$ plots in the temperature range $12^{\circ}$C - $60^{\circ}$C are studied at several drug concentrations at different $q$ values. For very low drug concentrations,  $C(\tau)$ changes from a stretched exponential to a non-exponential form at around $23^{\circ}$C, the CMT of pure F127 solutions. When drugs are added to the micellar solutions above a thresold concentration $C_{t}$,  a rather abrupt decrease in the solution CMT is noticed, and the auto-correlation function is always of stretched exponential form in the experimental temperature range.  In Table 1, the octanol-water partition coefficients (log [$P_{oct/water}$]) \cite{jones_partition}, which are measures of the hydrophobicity of  the three drugs used here, are listed alongside our estimates of $C_{t}$. It is seen that an increase in the partition coefficient reduces $C_{t}$, thereby favoring the formation of micelles at lower concentrations.

 %\begin{table}

%\begin{center}

%\begin{tabular}{|r|r|}
%	\hline
%$Drugs$ &  $C_{critical}$   \\
%	\hline
%Ibuprofen &    0.03 wt\%    \\
%Erythromycin & 0.06 wt\%  \\
%Aspirin &      0.3 wt\%  \\
	%\hline
%\end{tabular}\\
%\caption{Critical concentration for CMT change}
%\label{Table 1}

%\end{center}

%\end{table}
 SAXS measurements were also performed to confirm the lowering of the CMT of an Ibuprofen-encapsulated F127 solution above $C_{t}$. The data is shown in the inset (b) of Fig. 4, where the scattered X-ray intensity $I(q)$ is plotted \textit{vs.} $q$ for $60^{\circ}$C (circle-line) and $4^{\circ}$C (solid line). At $60^{\circ}$ C, a peak in $I(q)$ at $q$ = 0.004 1/nm is seen, which indicates the presence of micellar structures with diameters of $2\pi/q$ = 15.7$\pm$0.1 nm in solution. At $4^{\circ}$C, the peak completely disappears, confirming the absence of spherical micelles. Combining the X-ray and DLS measurements, it is concluded that the CMT of the Ibuprofen-encapsulated F127 micellar solutions lies somewhere in the range between $4^{\circ}$C and $12^{\circ}$C.  This confirms our earlier observation that drug encapsulation stabilizes F127 micelles over a broader temperature range.\\
%\begin{figure}
%\begin{center}
%\includegraphics[width=3.5in]{XRD_data_25BF.pdf}
%\caption{I(q) vs q plot for the SAXS study of 0.05 g/cc F127 with 0.25 wt\% Ibuprofen addition at $60^{\circ}$ C(blue), $4^{\circ}$ C (red) C.\\}
%\label{FIG 6}
%\end{center}
%\end{figure}
 \indent Aspirin and Erythromycin were also added to 5.26 wt\% F127 solutions above their respective $C_{t}$ values and DLS measurements were performed. Similar to the data of Fig. 4, the $ C(\tau)$ {\it vs.} $\tau$ plots show fits to stretched exponential forms in the temperature range 12$^{\circ}$C - 60$^{\circ}$C (representative data shown in Figs. S2(a) - S2(d) of Supporting Information). For every experiment, the relaxation time $\tau_{R}$ is obtained at each $q$. It is seen that $1/<\tau_{R}>$ varies linearly  with $q^{2}$ with $1/<\tau_{R}>~ \rightarrow 0$ as $\textit{q}$ $\rightarrow 0$ in all cases (insets of Figs. S2(a)- S2(d)). The same trends are repeated when drugs are incorporated in more dilute F127 solutions. Data for a 2.04 wt\% F127 solution are displayed in Figs. S2(e) and S2(f) of Supporting Information. The slopes of the linear fits to the $1/<\tau_{R}>$ \textit{vs}. $q^{2}$ data yield the translational diffusion coefficients D and the average hydrodynamic radii $<R_{H}>$ of the micelles (Fig. S2 and Table S1 of Supporting Information). \\
 
 In Fig. 5, the variations of $<R_{H}>$ of 5.26 wt\% F127 solutions, without drugs and when encapsulating 0.1 wt.\%  Ibuprofen, Aspirin and Erythromycin, respectively, are plotted \textit{vs.} temperature. For all the three drugs, $<R_{H}>$ decreases with increase in temperature, with the increase becoming sharper at temperatures below $30^{\circ}$C. Polydispersities also increase substantially when the temperature is lowered and when drugs are incorporated in the micellar cores. The inset of Fig. 5 displays the distributions of relaxation times for all the samples at 40$^{\circ}$C. A broader distribution function, and therefore a larger full width at half maximum (FWHM) value of the distribution, indicates higher polydispersity. Micellar polydispersities of all the samples were estimated using the protocol detailed in Part B of the `Analysis and Calculations' section of Supporting Information. It is seen that Aspirin, which is the least hydrophobic drug used, forms the largest and most polydisperse micelles at this temperature, while Ibuprofen, the most hydrophobic drug used here, forms  smaller, more compact micelles.  The same trends are repeated at 60$^{\circ}$C (Fig. S3 of Supporting Information). At 25$^{\circ}$C, however, Erythromycin incorporation causes the formation of the largest and most polydisperse micelles (Fig. S4 of Supporting Information). Greater drug hydrophobicity therefore ensures more compact packing of the drugs within the micellar cores at temperatures $\ge$ 40$^{\circ}$C. This trend is repeated even when F127 concentration is reduced (Figs. S5 and S6 of Supporting Information).

The morphologies of the micelles are therefore extremely sensitive to the molecular architectures and hydrophobicities of their constituents. The average hydrodynamic radii of the micelles $<R_{H}>$ at various temperatures and the polydispersity indices (PDIs) of these micelles when they encapsulate different quantities of drug solutes are plotted in Figure S7 of Supporting Information as the main figure and inset, respectively. The incorporation of drugs increases the radii of the micelles (Figs. 5 and Fig. S7), in agreement with previous experiments involving the encapsulation of drug molecules in Pluronic block copolymer micelles \cite{sharma_ijp,sharma_collsurf,foster_lang}.  It is important to note here that the small sizes of the drug-encapsulated micelles estimated here indicate that they may be sterilized simply by a filtration process and that the micelles may be easily delivered to the blood stream through intravenous injection.   It is, however, not possible to extract any systematic correlations between average micellar sizes and incorporated drug concentrations due to the very high micellar polydispersities. The data for the lower F127 concentrations also display the same trends (Fig. S8 of Supporting Information). The thermo-reversibility of the aggregation process has also been verified. The $<R_{H}>$  data estimated from DLS experiments for increasing and decreasing temperature cycles when 0.1 wt\% Ibuprofen is incorporated in a 5.26 wt\% F127 solution is plotted in Fig. S9 of Supporting Information. \\
       
%\indent
%\begin{figure}
%\begin{center}
%\includegraphics[width=3.5in]{02DLS_1.pdf}
%\caption{(a)The intensity auto-correlation functions C($\tau$) plotted on a log-linear scale for 0.02 g/cc F127 sample at $60^{\circ}$ C with addition of  0.25 wt\%  Ibuprofen, with the corresponding stretched exponential fits shown by solid lines. In the inset, the plot of  $\tau_{R}$  versus the square of the wavevector $q^2$  show a linear fit to the data; (b)Variation of hydrodynamic radius $R_{H}$ vs temperature  for 0.02 g/cc F127 solution with the encapsulation of different concentration of  Ibuprofen: 0 wt$\%(\square$), 0.1 wt$\%(\circ$), 0.25 wt$\%(\triangle$).\\}
%\label{FIG 8}
%\end{center}
%\end{figure}
\indent To confirm the increased hydration of the micellar core due to decrease in solution temperature, fluorescence spectra of the samples were recorded using aromatic pyrene molecules as the fluorescent probe. Pyrene molecules, being hydrophobic, stay encapsulated in the PPO cores of the micelles. The samples were excited at a wavelength of 339 nm with a bandwidth of 1 nm. The emission spectra, recorded at $40^{\circ}$C and $8^{\circ}$C, are shown in Fig. 6(a) (solid line and circle-line, respectively). In both these spectra, the first peak, $I_{1}$, occurs  at around 372 nm and the third peak, $I_{3}$, occurs at around 383 nm.\\
\indent  Earlier studies have shown that the intensity ratio between the first and third emission peaks of the  pyrene spectrum ($I_{1}/I_{3}$) depends upon the dipole moment of the solvent \cite{naka_1,naka_2,kalyan_jacs}.  A significant enhancement in the intensity of the 0-0 vibronic band (which results in the $I_{1}$ peak) in the presence of a polar solvent has been reported. The $I_{3}$ peak, in contrast, is solvent-insensitive. A transition from a non-polar to a polar environment therefore results in a significant increase of the $I_{1}/I_{3}$ ratio in the pyrene emission spectrum. In Fig. 6(b), the ratios $I_{1}/I_{3}$ are plotted \textit{vs.} temperature for different concentrations of drugs added to F127 solutions. For every sample, the ratio increases with decrease in temperature. This indicates an increase in the aqueous content of the micellar core region at lower temperatures. Increased solvent penetration into the micellar cores should result in looser packing of molecules in the core and an increase in the values of $<R_{H}>$ at lower temperatures. This has already been reported in Figs. 5 and S7. For the samples with CMTs above $12^{\circ}$C (F127 micelles encapsulating 0 wt\% Ibuprofen and 0.05 wt\% Erythromycin, respectively), $I_{1}/I_{3}$ shows a very high value of around 1.7 at $8^{\circ}$ C which indicates that the pyrene molecules are in a fully aqueous environment \cite{kalyan_jacs}. The high values of $I_{1}/I_{3}$ seen in our measurements for low drug encapsulation (Fig. 6(b)) indicate that the pyrene molecules are released into the aqueous environment at $8^{\circ}$C due to the breakup of the micelles into unimers. For the other two samples with  higher drug encapsulation, the increase in $I_{1}/I_{3}$  at lower temperature is much less, which indicates a more limited amount of sovent penetration in the micellar cores.  \\ 

%\begin{figure}
%\begin{center}
%\includegraphics[width=3.5in]{fluro2.pdf}
%\caption{ The normalised intensity auto-correlation functions C($\tau$) plotted on a log-linear scale for 0.05g/cc F127 sample with 0.5 wt\% at $60^{\circ}$C ($\circ$) and $15^{\circ}C$ ($\square$) at pH 11.36. In the inset, $I_{1}/I_{3}$ ratio vs temperature from the flueoscence measurement of 0.5 wt\% Ibuprofen added 0.05 g/cc F127 solution for pH value adjusted to 11.36.\\}
%\label{FIG 11}
%\end{center}
%\end{figure}
 The addition of the weakly acidic Ibuprofen molecules to non-ionic F127 solutions results in a monotonic decrease in the sample pH. For example, the addition of 0.5 wt\% Ibuprofen to a 5.26 wt\% F127 solution decreases the solution pH from 7 to 4.65. Small amounts of NaOH were added to change the solution pH and systematic DLS measurements were next performed.  The normalized intensity auto-correlation functions $C(\tau)$, obtained in DLS measurements for 5.26 wt\% F127 solutions containing 0.5 wt\% Ibuprofen at $40^{\circ}$C for different solution pH values, fit to stretched exponential functions (circles in Fig. 7(a) represent data at pH = 11.36). When the temperature is decreased to 15$^{\circ}$C, the micelles dissociate into free unimers and $C(\tau)$ [squares in Fig. 7(a)] does not fit to a stretched exponential form. This is consistent with the high $I_{1}/I_{3}$ ratios obtained in fluorescence measurements on the same sample [inset of Fig. 7(a)]. 

Inset of Fig. 7(b) shows that as solution pH is changed between 4.65 and 11.36,  $1/<\tau_{R}>$, extracted from DLS data acquired  40$^{\circ}$C, shows a power law dependence on $q$: $1/<\tau_{R}>~\approx~q^{\alpha}$. Here, the exponent $\alpha$ gives information about the micellar dynamics. Fig. 7(b) plots the values of $\alpha$ and $1/<\tau_{R}>$ {\it vs.} solution pH. At the low pH of 4.65, the power law fit gives $\alpha$ $\approx$ 2, suggesting that the system consists of homogeneous, spherical, diffusive structures, consistent with the results displayed in Fig. 4. With an increase in solution pH to 6.1, $\alpha$ increases to 2.57, indicating the presence of heterogeneous micellar structures in solution. This increase in $\alpha$ is accompanied by a simultaneous decrease in $1/<\tau_{R}>$, which indicates the presence of bigger aggregates. When solution pH is increased to 11.36, a decrease of $\alpha$ to 2.19 and an increase in $1/<\tau_{R}>$ are observed. \\

\noindent As changing pH should not affect the non-ionic F127 micelles, the observed changes are brought about by the Ibuprofen molecules present in solution. With increase in solution pH, the Ibuprofen molecules in the micellar core are ionized \cite{foster_lang2,alexander_lang}. The repulsive interactions within the core lead to the formation of bigger, more anisotropic micelles at pH = 6.1.  When the pH of the system is further increased to 11.36, the Ibuprofen molecules, which are fully ionized, are released from the micellar cores to the aqueous medium due to their enhanced water solubility. The encapsulation of drug molecules  is not favored under these conditions and  pure F127 micelles coexist in solution with free drug molecules. This results in the observed decrease in $\alpha$.  It is to be noted here that the average hydrodynamic radii $<R_{H}>$ of the micelles estimated in DLS measurement is 10.4 nm at T = $40^{\circ}$C and pH = 11.36, almost identical to the value extracted when no drugs are added (Figs. 5 and S7, Table S1).  Furthermore, by observing changes in the shapes of $C(\tau)$ from stretched exponential to non-exponential as temperature is decreased, the CMT of the sample at pH = 11.36 is estimated to be  $24^{\circ}$C, the value expected for pure F127 micellar solutions. This observation is similar to the findings of Scherlund \textit{et al.}, who observed that if the active ingredients (drugs) are mostly in ionized form at a certain pH, the CMT of the F127 solution is almost the same as that for the pure sample \cite{scher_ijp}. The release of drugs at high pH was previously also reported for Flurbiprofen-encapsulated P103 solutions \cite{alexander_lang}. 
 
 \section{Conclusion}
  In this work, three different drugs, Ibuprofen, Aspirin and Erythromycin, are encapsulated in aqueous F127 micellar solutions. Cryo-SEM imaging shows the presence of globular structures even after drug-encapsulation.  The drug molecules, when added above a specific threshold concentration $C_{t}$, enhance the stability of the micellar phase over a broad temperature range. DLS experiments are performed on the drug-encapsulated micellar systems to determine changes in hydrodynamic radii and micellar polydispersities. The average  hydrodynamic radii of the micelles calculated here are seen to increase upon drug incorporation, with the polydispersity being inversely correlated to the drug hydrophobicity at the higher temperatures. The hydrodynamic radii of the micelles increase with decrease in temperature. This is consistent with data presented in \cite{sharma_ijp,sharma_collsurf}.  The increase in micellar sizes observed when temperature is lowered is accompanied by an increase in the hydration of the micellar core and is verified by pyrene fluorescence spectra measurements. Increasing the temperature excludes the solvent from the micellar core due to enhanced core hydrophobicity and results in more compact micelles. The drug-encapsulated micelles also show a strong pH dependence. When pH increases from 4.65 to 6.1, the ionization of drug molecules leads to the formation of larger micelles. Increasing the pH to 11.36 triggers the release of drug molecules in the solvent. Drug-incorporated F127 micelles therfore have enormous potential as drug carriers in the area of nanomedicine \cite{moghimi_nanotech}. However, before these formulations are made available commercially, several clinical and scientific issues relating to their toxicity and stability need to be addressed.

 \section{Acknowledgment} The authors thank A. Pal and V. A. Raghunathan for their help with the SAXS measurements, P. Rose and  R. Philip for their help with  fluorescence measurements and D. Chelvan for his assistance with the cryo-SEM measurements. The authors are also grateful to D. Saha for his help in computing micellar polydispersities from the DLS data. \\

{\bf Supporting Information available:} This material is available free of charge via the Internet.

\newpage 
\begin{center}
    \begin{table}
\label{TABLE 1}
\caption{The octanol-water partition functions log [$P_{oct/water}$] of the three drugs used here and the $C_{t}$ values obtained  from the DLS data are tabulated below. \\ }
\begin{tabular}{| l | l | l |}
  \hline
    Name of drug & log [$P_{oct/water}$] & $C_{t}$ \\ \hline
   Ibuprofen & 3.5 & 0.03 wt.\%\\ \hline
     Aspirin & 1.19 & 0.3 wt.\%\\\hline
Erythromycin & 3.06 & 0.05 wt.\%\\\hline
\end{tabular}
 \end{table}
\end{center}

\begin{figure}
\begin{center}
\includegraphics[width=3.25in]{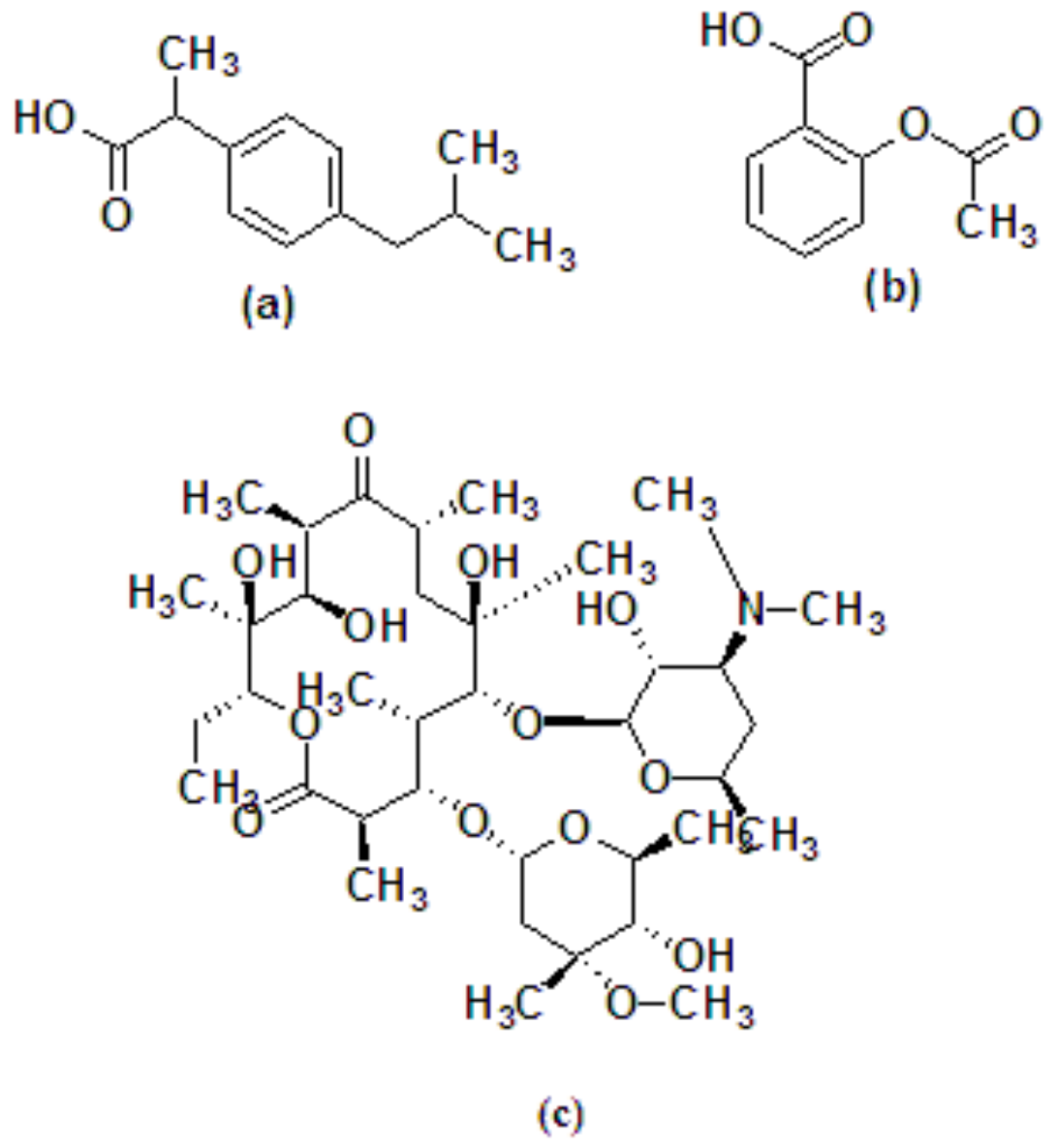}
\caption{Chemical structures of (a) Ibuprofen, (b) Aspirin and (c) Erythromycin.}
\label{FIG 1}
\end{center}
\end{figure}

\begin{figure}
\begin{center}
\includegraphics[width=3.25in]{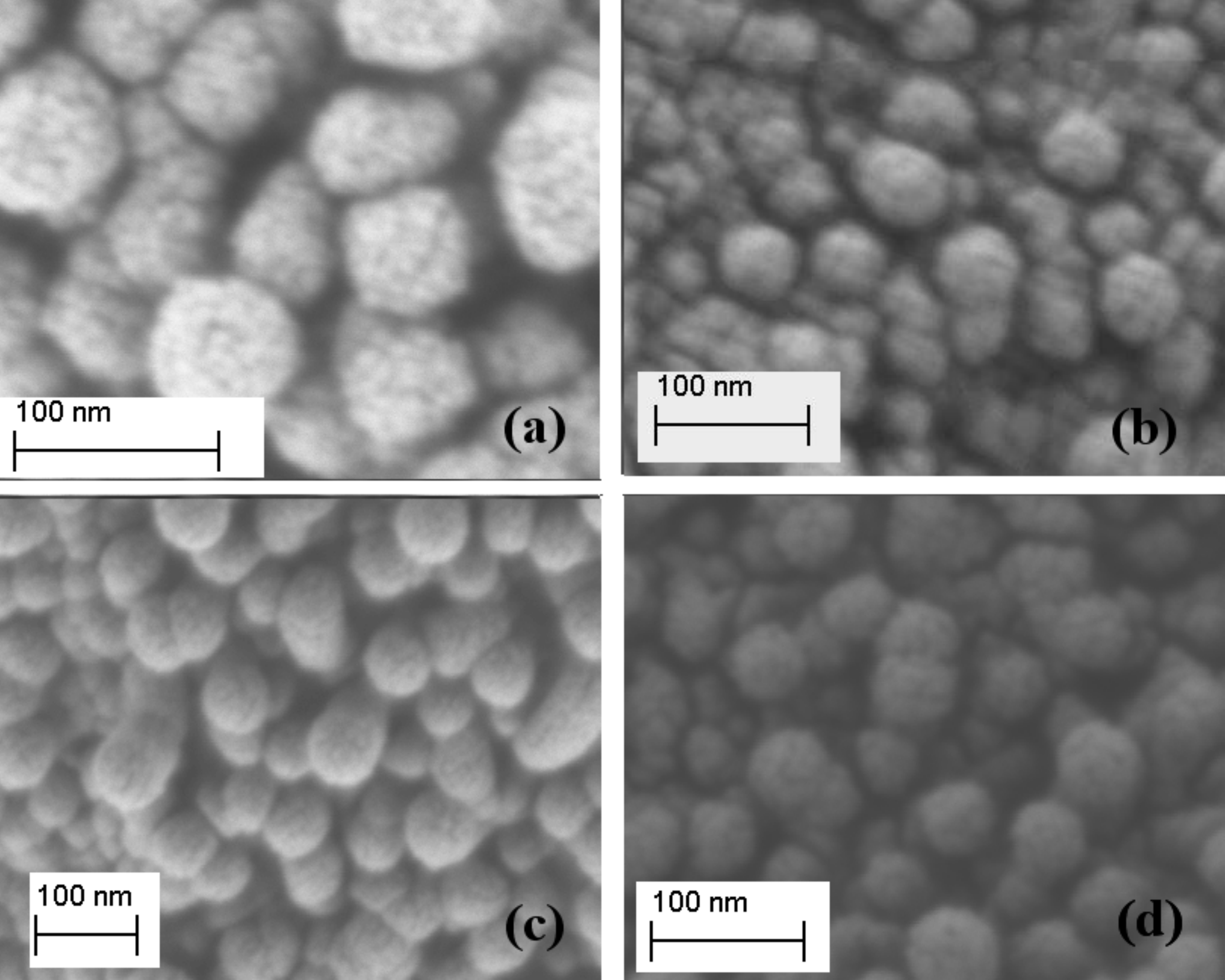}
\caption{Cryo-SEM images of 5.26 wt\%  F127 micelles (a) with no encapsulated drugs, and when (b) 0.1 wt\% Ibuprofen, (c) 0.1 wt\% Aspirin and (d) 0.1 wt\% Erythromycin are incorporated in the micellar cores.   \\}
\label{FIG 2}
\end{center}
\end{figure}

%\begin{figure}
%\begin{center}
%\includegraphics[width=3.25in]{final_fig3.pdf}
%\caption{Size distributions, derived from the cryo-SEM images of micellar structures, for 5.26 wt\% F127 samples (a) without drugs, %(b) with 0.25 wt\% Ibuprofen, (c) with 0.5 wt\% Aspirin and (d) with 0.05 wt\% Erythromycin.   \\}
%\label{FIG 3}
%\end{center}
%\end{figure}

\begin{figure}
\begin{center}
\includegraphics[width=3.25in]{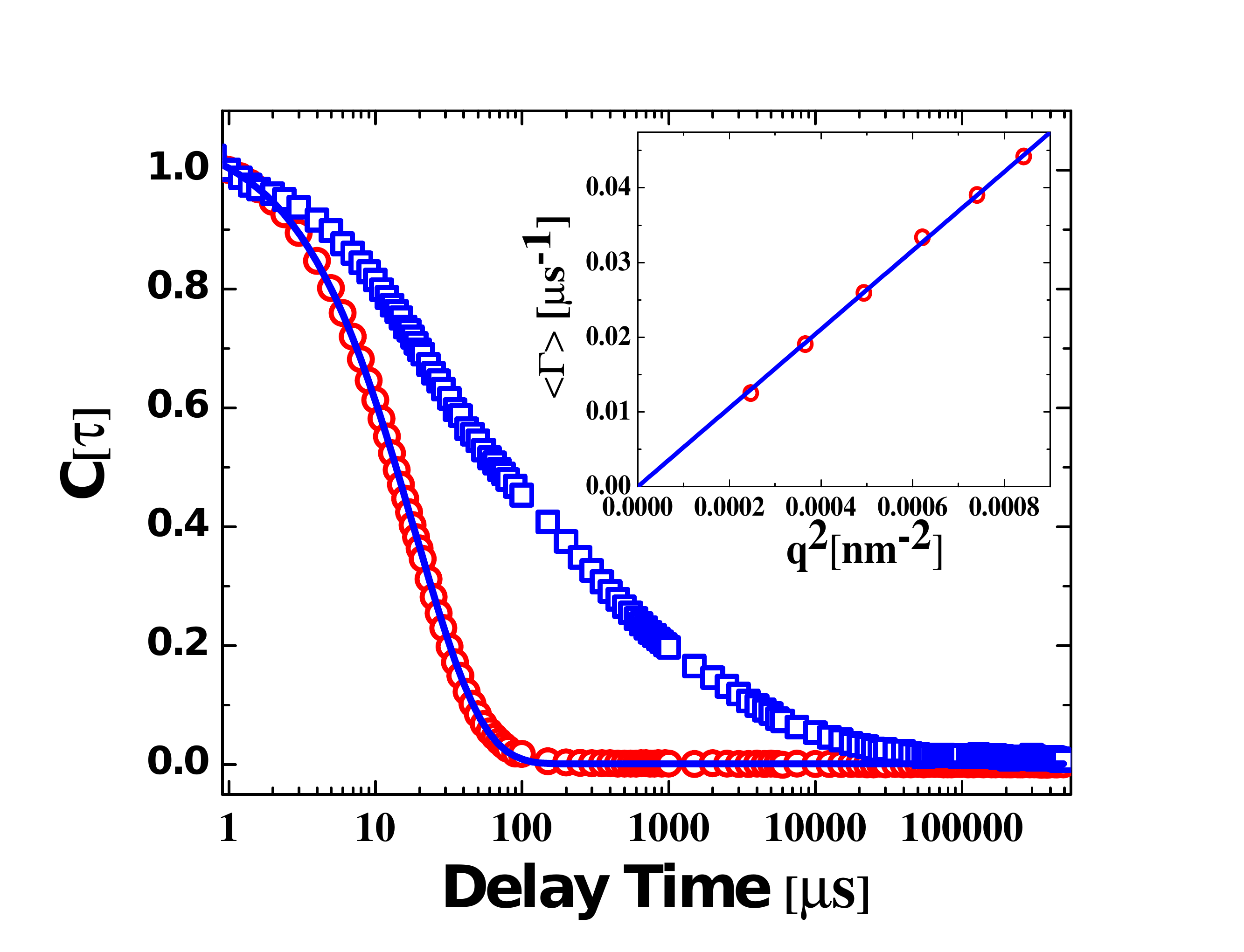}
\caption{ The normalized intensity auto-correlation functions $C(\tau)$ plotted for a pure 5.26 wt\% F127 sample at 20$^{\circ}$C (circles), 25$^{\circ}$C (up-triangles), 40$^{\circ}$C (down-triangles) and $60^{\circ}$C (squares). The stretched exponential fits to the data at 60$^{\circ}$C, 40$^{\circ}$C and 25$^{\circ}$C are shown by solid lines. The linear fits to $1/<\tau_{R}> $ \textit{vs.} $q^{2}$, obtained at $60^{\circ}$C, 40$^{\circ}$C and 25$^{\circ}$C, are shown in the inset (solid lines).\\}
\label{FIG 3}
\end{center}
\end{figure}

\begin{figure}
\begin{center}
\includegraphics[width=3.25in]{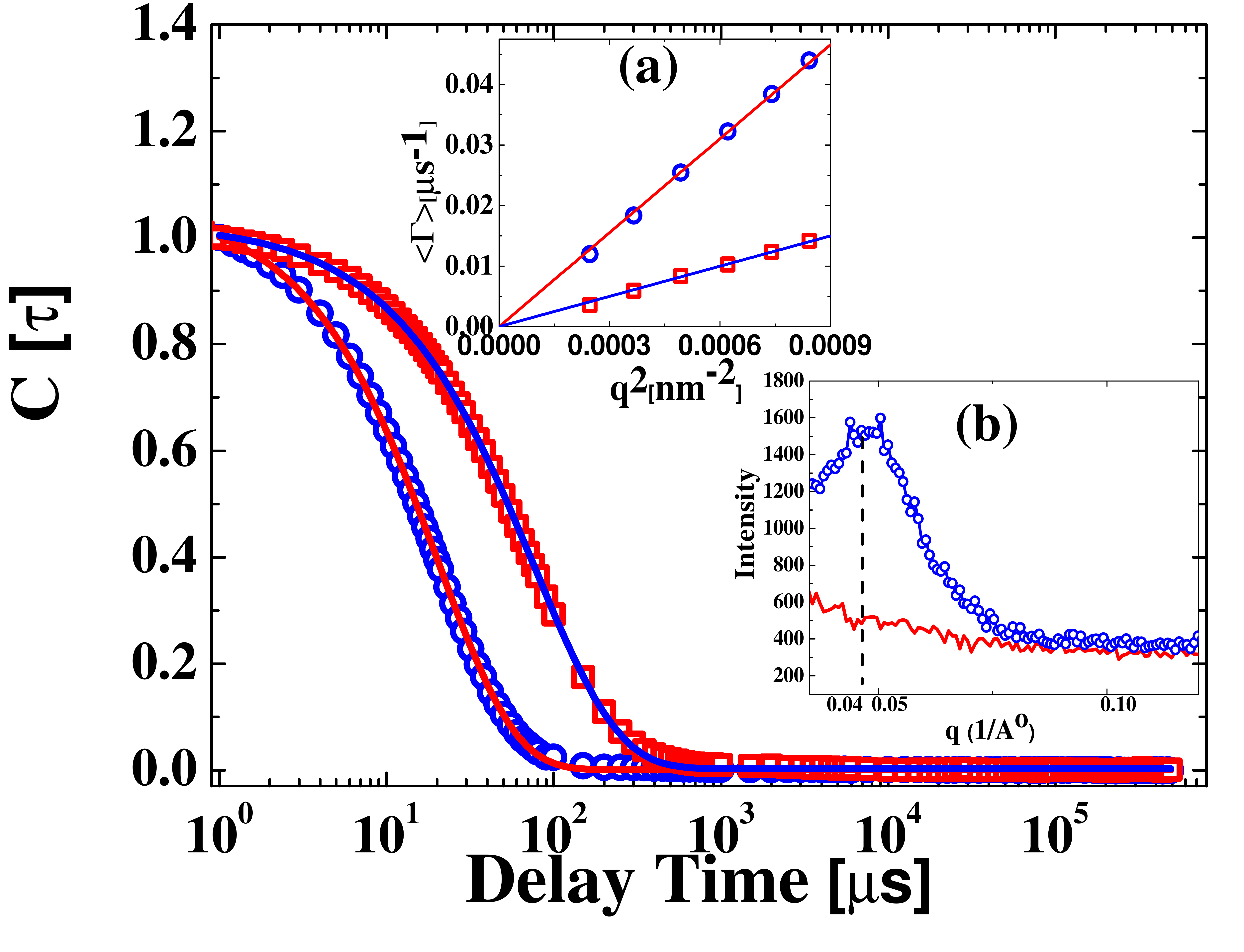}
\caption{ The normalized intensity auto-correlation functions $C(\tau)$ plotted for 5.26 wt\% F127 solutions with 0.1 wt\% Ibuprofen at 20$^{\circ}$C (circles), 25$^{\circ}$C (up-triangles), 40$^{\circ}$C (down-triangles) and 60$^{\circ}$C (squares), respectively. The stretched exponential fits are shown by solid lines. The inset (a) shows that $1/<\tau_{R}>$  varies linearly with $q^2$ for all the samples. (b) SAXS data showing $I(q)$ \textit{vs.} $q$ for a  5.26 wt\% F127 solution incorporating 0.25 wt\% Ibuprofen is plotted at $60^{\circ}$C (circle-line) and $4^{\circ}$C (solid line).\\}
\label{FIG 4}
\end{center}
\end{figure}

%\begin{figure}
%\begin{center}
%\includegraphics[width=3.25in]{final_fig6.pdf}
%\caption{The normalized intensity auto-correlation functions C($\tau$) plotted at $60^{\circ}$C for 0.05 g/cc F127 sample with %addition of (a) 0.5 wt\% Aspirin and (b) 0.05 wt\% Erythromycin. In (c), DLS data for 0.02 g/cc F127 sample with 0.25 wt\% %Ibuprofen at $60^{\circ}$C is plotted. Stretched exponential fits are shown by solid lines. In the insets,  $1/<\tau_{R}>$  %\textit{vs.} $q^2$ plots show linear fits for all three samples.\\}
%\label{FIG 5}
%\end{center}
%\end{figure}

\begin{figure}
\begin{center}
\includegraphics[width=3.25in]{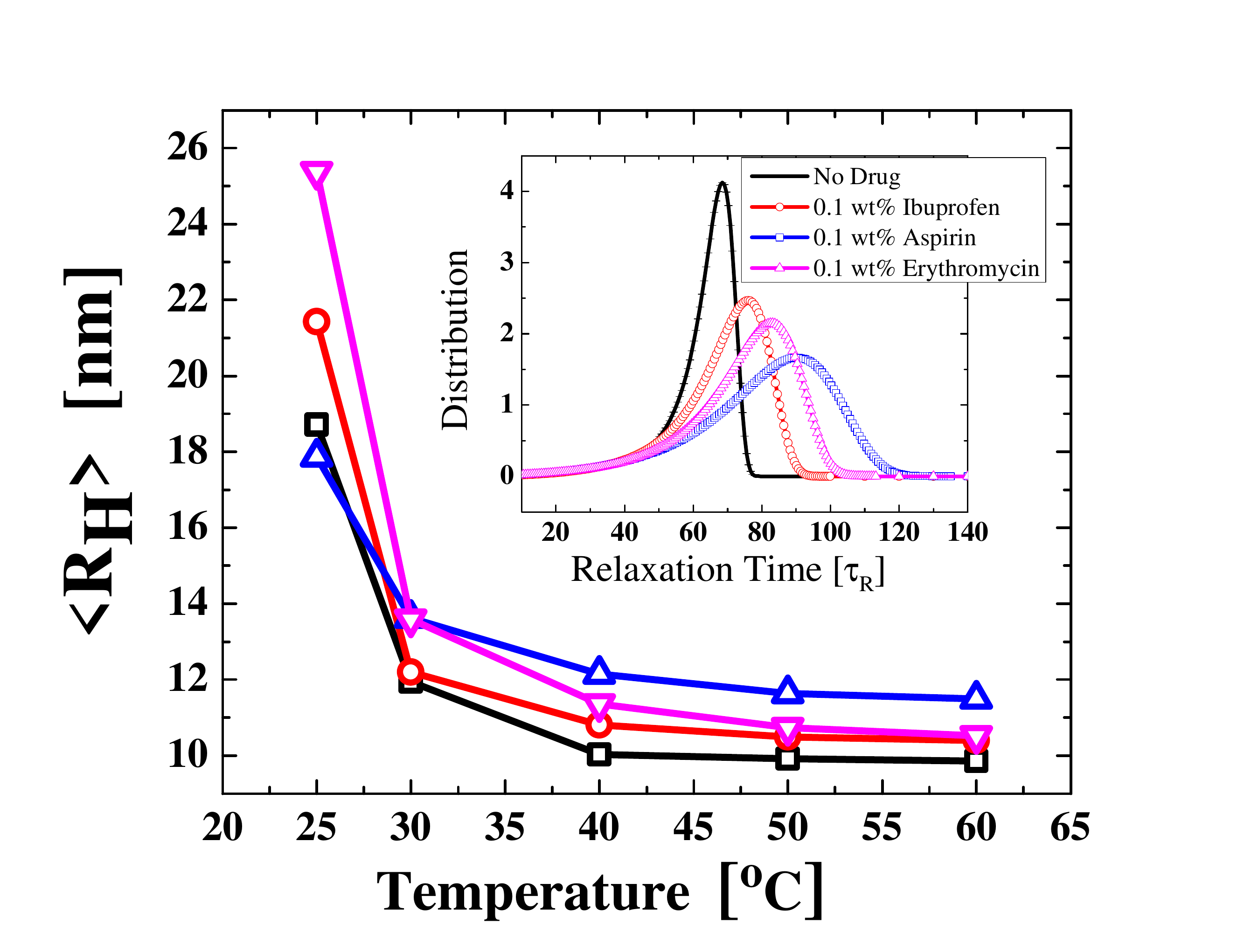}
\caption{Variation of $<R_{H}>$ with temperature  for a 5.26 wt\% F127 solution without drugs (squares) and after the encapsulation of 0.1 wt\% Ibuprofen (circles), 0.1 wt\% Aspirin (up-triangles) and 0.1 wt\% Erythromycin (down-triangles). Inset shows the distributions of relaxation times for F127 micellar solutions with no drugs (solid line), 0.1wt\% Ibuprofen (circle-line), 0.1 wt\% Aspirin (square-line) and 0.1 wt\% Erythromycin (triangle-line) at 40$^{\circ}$C.\\}
\label{FIG 5}
\end{center}
\end{figure}
\begin{figure}
\begin{center}
\includegraphics[width=3.25in]{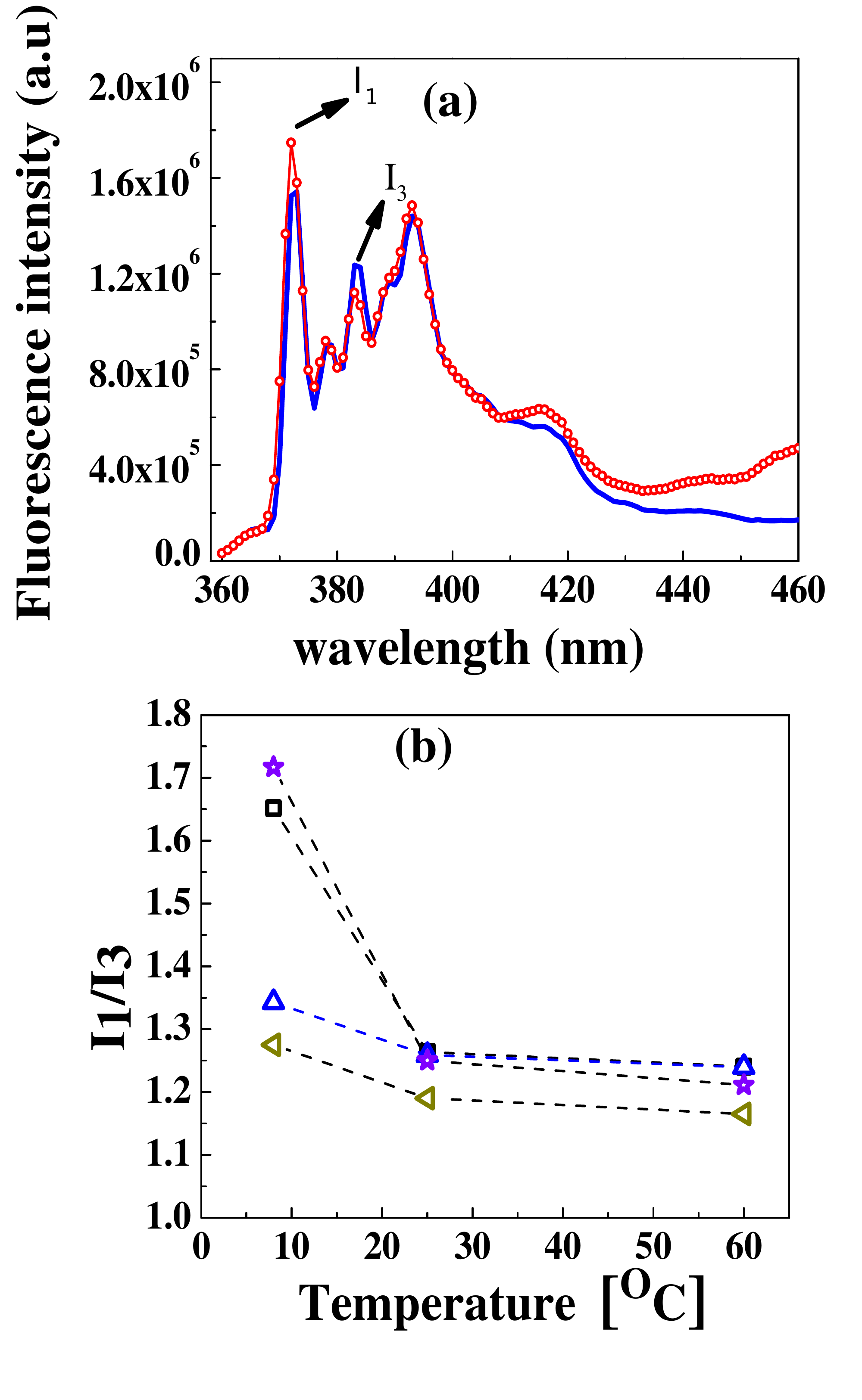}
\caption{(a) Pyrene emission spectra for 5.26 wt\%  F127 solutions at $40^{\circ}$C (solid line) and $8^{\circ}$C (circle-line).  Variations of $I_{1}/I_{3}$ with temperature when 0 wt\% Ibuprofen (squares), 0.2 w\% Ibuprofen (up-triangles), 0.5wt\% Aspirin (left-triangles) and 0.05 wt\% Erythromycin (stars) are incorporated in F127 solutions are plotted in (b). \\}
\label{FIG 6}
\end{center}
\end{figure}

\begin{figure}
\begin{center}
\includegraphics[width=3.25in]{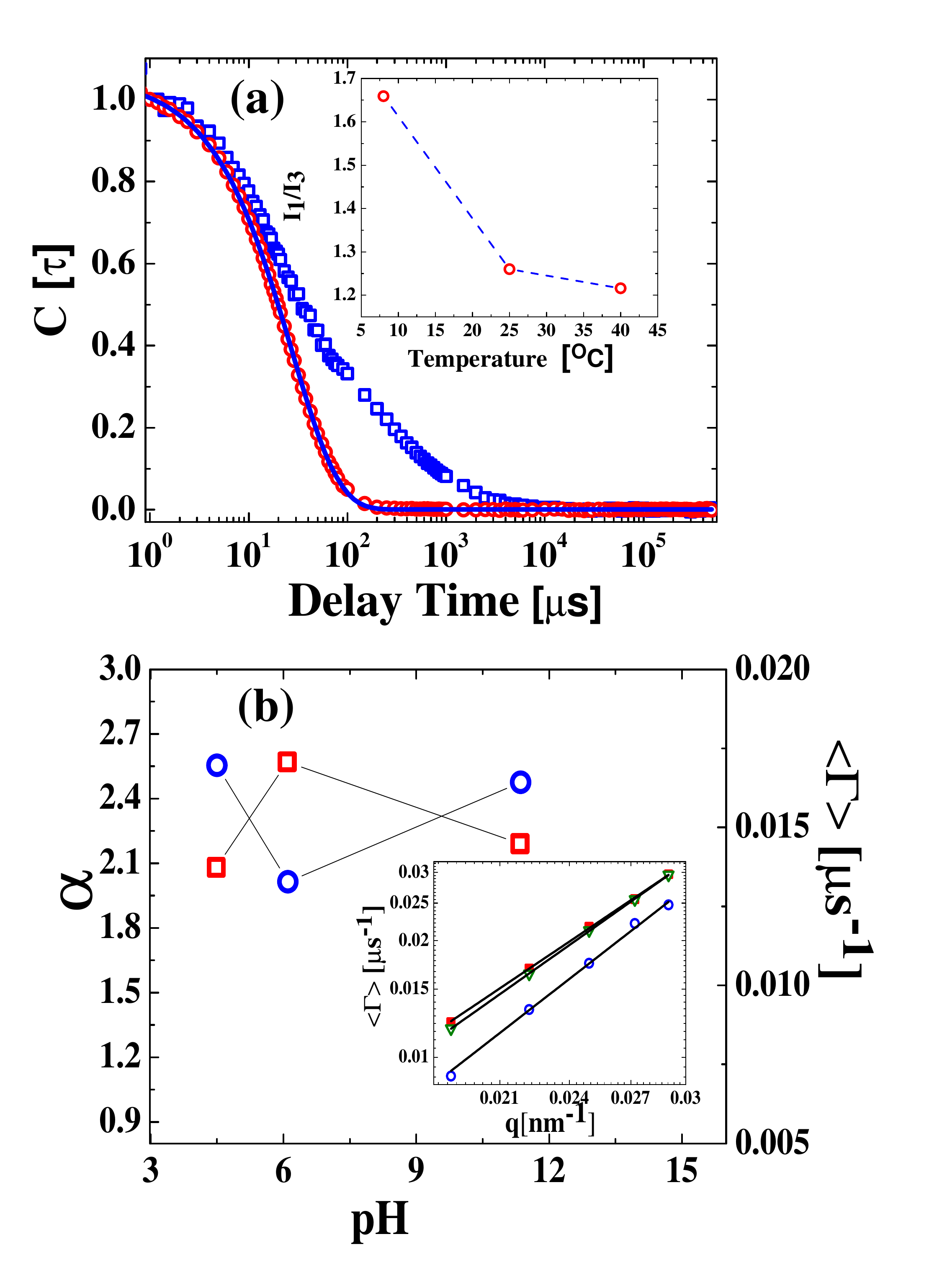}
\caption{(a) The normalized intensity auto-correlation functions $C(\tau)$ at $\theta$ = $90^{\circ}$ for 5.26 wt\% F127 solutions containing 0.5 wt\% Ibuprofen at $40^{\circ}$C (circles) and $15^{\circ}$C (squares) respectively, when the solution pH is adjusted to 11.36, are shown. $I_{1}/I_{3}$ \textit{vs.} temperature is plotted in the inset of (a). The values of $\alpha$ (squares) and $1/<\tau_{R}>$ (circles) at  40$^{\circ}$C are plotted \textit{vs.} solution pH in (b). Inset of (b) shows the plots of $1/<\tau_{R}>$ \textit{vs.} $q$ and the corresponding fits (solid lines) to $1/\tau_{R} \approx q^{\alpha}$ at $40^{\circ}$C for different pH values: 4.5 (squares), 6.1 (circles), 11.36 (triangles).}
\label{FIG 7}
\end{center}
\end{figure}

\pagebreak
%\lhead{TOC Graphic}
%\rhead{\today}
\begin{figure}
\begin{center}
\includegraphics[width=5in]{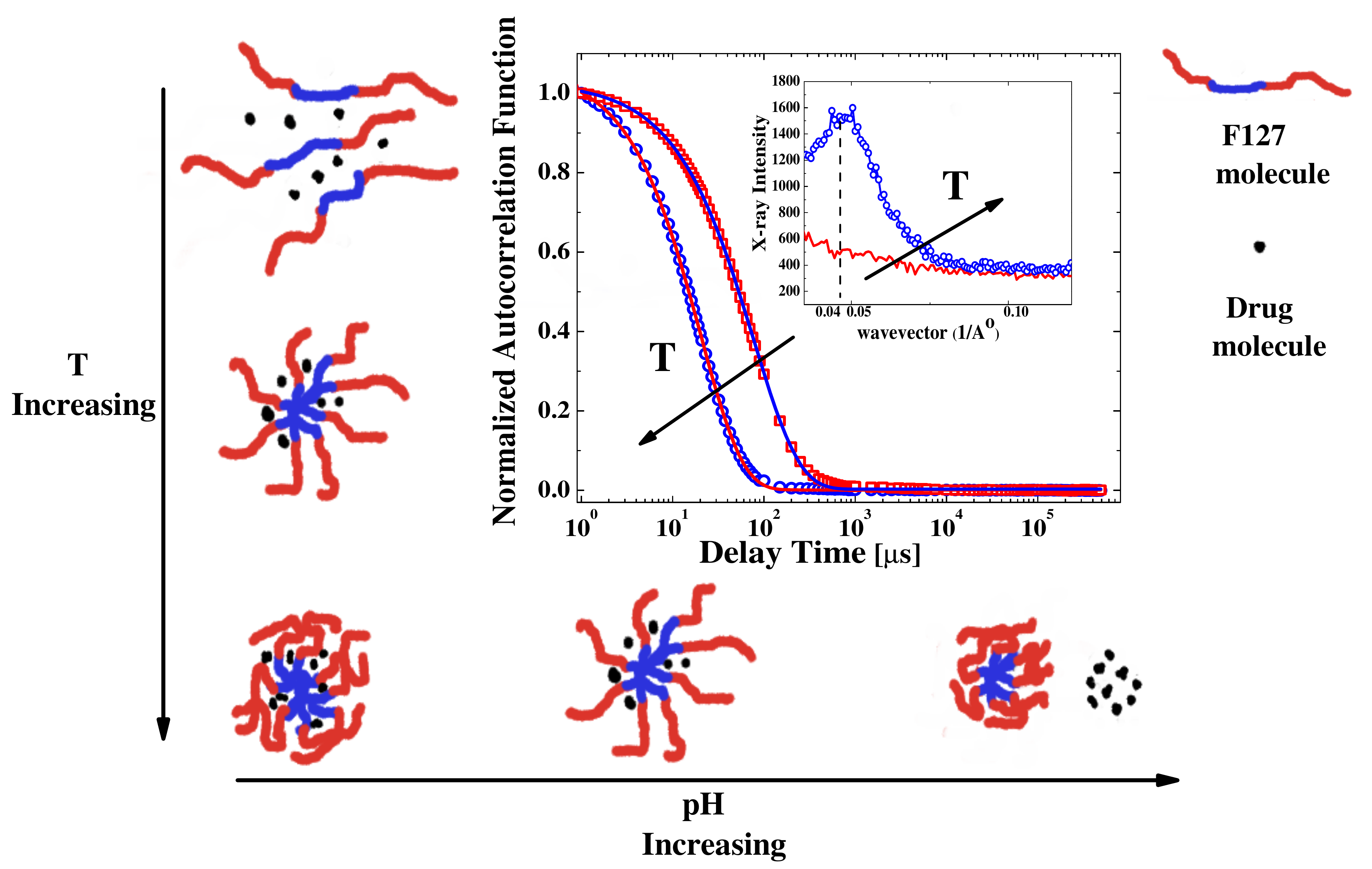}
\caption{TOC Graphic}
\end{center}
\end{figure}
\end{document}